\documentclass[aps,prc,twocolumn,groupedaddress,showpacs]{revtex4-1}

\usepackage{graphicx}
\usepackage{bm}

\begin{document}


\title{Further evidence for three-nucleon spin-orbit interaction
in isotope shifts of $Z=\mathrm{magic}$ nuclei}


\author{H. Nakada}
\email[E-mail:\,\,]{nakada@faculty.chiba-u.jp}
\affiliation{Department of Physics, Graduate School of Science,
 Chiba University,\\
Yayoi-cho 1-33, Inage, Chiba 263-8522, Japan\\}
\affiliation{Institut de Physique Nucl\'{e}aire de Lyon,
 Universit\'{e} de Lyon 1,
F-69622 Villeurbanne, France}


\date{\today}

\begin{abstract}
It was pointed out [Phys. Rev. C \textbf{91}, 021302(R)]
that the isotope shifts of the Pb nuclei, the kink at $N=126$ in particular,
can be well described by the Hartree-Fock-Bogolyubov calculations
if a density-dependent LS interaction
derived from the $3N$ interaction is incorporated.
Effects of the density-dependence in the LS channel on the isotope shifts
are extensively investigated for the Ca, Ni and Sn isotopes,
using the semi-realistic M3Y-P6 interaction
and its LS modified variant M3Y-P6a, as in the Pb case.
It is found that almost equal charge radii between $^{40}$Ca and $^{48}$Ca
are reproduced,
as well as the isotope shifts in a long chain of the Sn nuclei,
owing to the density-dependence in the LS channel.
A kink is predicted at $N=82$ for the isotope shifts of the Sn nuclei,
in clear contrast to the interactions without the density-dependence.
\end{abstract}


\maketitle



\section{Introduction\label{sec:intro}}
In the nuclear shell structure,
which is formed by a series of the nucleonic single-particle (s.p.) orbits
under the nuclear mean field (MF),
the spin-orbit ($\ell s$) splitting plays an essential role.
However, though known from the data,
size of the $\ell s$ splitting has been difficult to be accounted for
only by the two-nucleon ($2N$) interaction~\cite{ref:AB81}.
Recent development of the chiral effective-field theory
($\chi$EFT)~\cite{ref:EGM05}
indicates that the three-nucleon ($3N$) interaction gives 
significant density-dependence in an LS channel
when it is converted to an effective $2N$ interaction~\cite{ref:Kai03,ref:Koh12},
and that this may account for the missing part
of the $\ell s$ splitting~\cite{ref:Koh12,ref:Koh13}.

There have been precise measurements
on isotope shifts~\cite{ref:AHS87,ref:Ang04}.
As well as the $\ell s$ splitting,
they have supplied problems in nuclear structure theory
that have not been solved for decades.
As an important example,
the isotope shifts of the Pb nuclei show a conspicuous kink at $N=126$
when plotted as a function of the neutron number $N$.
This kink was almost reproduced in a relativistic MF approach~\cite{ref:SLR94},
though not in non-relativistic approaches with the Skyrme interactions.
This difference has been recognized to originate from the isospin-dependence
of the LS channels~\cite{ref:SLKR95},
and has lead to extension
of the Skyrme energy density functional~\cite{ref:RF95}.
The $n1g_{9/2}$ and $n0i_{11/2}$ s.p. levels are relevant to the kink at $N=126$,
which are nearly degenerate in the relativistic model
and in the non-relativistic model with modified LS channels
of Ref.~\cite{ref:RF95}.
However, such degeneracy is not observed in $^{209}$Pb~\cite{ref:TI}.
The author, together with a coauthor,
pointed out that the density-dependence in the LS channel
enables us to reproduce the isotope shifts of the Pb nuclei fairly well,
without degeneracy between the $n1g_{9/2}$
and $n0i_{11/2}$ orbitals~\cite{ref:NI15}.
This effect on the isotope shifts provides us with evidence
for the $3N$ LS interaction
that is practically independent of the $\ell s$ splitting,
suggesting a possibility that the two problems,
origin of the $\ell s$ splitting and the isotope shifts in Pb,
are simultaneously solved by the $3N$ LS interaction.
In this paper effects of the density-dependent LS channel
are extensively examined for the isotope shifts
of other $Z=\mathrm{magic}$ nuclei.

Whereas a density-dependent LS interaction was argued in Ref.~\cite{ref:PF94},
motivated to account for the model-dependence
of the isotope shifts in the Pb nuclei,
primary origin of the difference between the relativistic
and the Skyrme approaches
has turned out to be the isospin-dependence of the LS channels.
Since then the density-dependence in the LS channel has been considered
only in a limited number of studies~\cite{ref:Gor15},
and its influence on physical quantities
other than energies has not been explored sufficiently.

\section{Framework\label{sec:Hamil}}
In the present work,
the spherical Hartree-Fock-Bogolyubov (HFB) calculations are implemented
for the Ca, Ni and Sn nuclei, as for Pb in Ref.~\cite{ref:NI15}.
Because the proton number $Z$ is magic in these nuclei,
it seems reasonable to investigate their systematic behavior
in the spherical HFB regime,
although there could be exceptions as will be mentioned below.
The computational method is an extensive application
of the Gaussian expansion method~\cite{ref:GEM},
and has been summarized in Ref.~\cite{ref:Nak08}.
The effective Hamiltonian is comprised of the nuclear, Coulomb
and center-of-mass (c.m.) parts: $H=H_N+V_C-H_\mathrm{c.m.}$.
The nuclear part $H_N$ is taken to be non-relativistic and isoscalar,
comprised of the central ($v^{(\mathrm{C})}$), LS ($v^{(\mathrm{LS})}$),
tensor ($v^{(\mathrm{TN})}$),
and the density-dependent central ($v^{(\mathrm{C}\rho)}$),
LS ($v^{(\mathrm{LS}\rho)}$) channels for the $2N$ interaction,
as well as of the kinetic energy term.
In the former papers~\cite{ref:Nak08b,ref:Nak10,ref:Nak13}
$v^{(\mathrm{C}\rho)}$ was denoted by $v^{(\mathrm{DD})}$.
For $v^{(\mathrm{LS}\rho)}$ the following form is assumed,
\begin{equation}
v_{ij}^{(\mathrm{LS}\rho)} = 2i\,D[\rho(\mathbf{R}_{ij})]\,
 \mathbf{p}_{ij}\times\delta(\mathbf{r}_{ij})\,\mathbf{p}_{ij}\cdot
 (\mathbf{s}_i+\mathbf{s}_j)\,, \label{eq:DDLS}
\end{equation}
where $\mathbf{r}_{ij}= \mathbf{r}_i - \mathbf{r}_j$,
$\mathbf{R}_{ij}=(\mathbf{r}_i+\mathbf{r}_j)/2$,
$\mathbf{p}_{ij}= (\mathbf{p}_i - \mathbf{p}_j)/2$,
$\rho(\mathbf{r})$ is the isoscalar nucleon density
and $\mathbf{s}_i$ is the spin operator,
with $i$ and $j$ being indices of constituent nucleons.
$V_C$ contains the exchange terms without approximation
and $H_\mathrm{c.m.}$ consists both of one- and two-body terms.

The Michigan-three-range-Yukawa (M3Y) interactions
were obtained from the $G$-matrix~\cite{ref:M3Y}.
By modifying the M3Y-Paris interaction~\cite{ref:M3Y-P} phenomenologically,
the semi-realistic M3Y-P$n$ interactions
have been developed~\cite{ref:Nak13,ref:Nak03},
whose density-independent terms $v^{(\mathrm{X})}$
($\mathrm{X}=\mathrm{C}$, $\mathrm{LS}$ and $\mathrm{TN}$)
are expressed by the Yukawa function.
Responsible for the shell structure,
the $\ell s$ splitting is significant to describing nuclear structure.
Except M3Y-P6a, $v^{(\mathrm{LS})}$ has been enhanced
from that of the original M3Y-Paris interaction,
to reproduce s.p. level sequence in the absence of $v^{(\mathrm{LS}\rho)}$.
With keeping the tensor channels $v^{(\mathrm{TN})}$ of the M3Y-Paris interaction,
$Z$- or $N$-dependence of the shell structure is obtained reasonably well,
as exemplified by the level inversion of $p0d_{3/2}$ and $p1s_{1/2}$
from $^{40}$Ca to $^{48}$Ca~\cite{ref:NSM13}.
Among several parameter-sets,
it has been found that M3Y-P6 gives prediction of magic numbers
compatible with most available experimental data,
in wide range of the nuclear chart
including unstable nuclei~\cite{ref:NS14}.
Here M3Y-P6 is taken as a yardstick
for investigating effects of the $3N$ LS interaction.
The values of the parameters~\cite{unit-corr} in M3Y-P6
have been given in Ref.~\cite{ref:Nak13}.

The $\chi$EFT derives density-dependence in the LS channel
as an effect of the $3N$ interaction~\cite{ref:Koh12,ref:Koh13},
which may complement the $2N$ LS interaction
with respect to the $\ell s$ splitting.
Based on this $\chi$EFT indication,
$v^{(\mathrm{LS}\rho)}$ has been introduced in Ref.~\cite{ref:NI15},
instead of enhancing $v^{(\mathrm{LS})}$,
which yields a variant of the M3Y-P6 interaction called M3Y-P6a.
However, since the currently available $\chi$EFT is not yet convergent
at $\rho\gtrsim\rho_0$, where $\rho_0$ denotes the saturation density,
the $\chi$EFT results have not completely been followed in quantitative respect.
The functional $D[\rho]$ is taken to be
\begin{equation}
D[\rho(\mathbf{r})] = -w_1\,\frac{\rho(\mathbf{r})}
 {1+d_1\rho(\mathbf{r})}\,. \label{eq:DinLS}
\end{equation}
The $d_1$ term of the denominator,
which has been fixed to be $1.0\,\mathrm{fm}^3$ in M3Y-P6a,
is employed to avoid instability
for increasing density,
whereas the results are not sensitive to $d_1$
as will be shown in Sec.~\ref{sec:result}.
Since the effective interactions have been more or less adjusted
to the observed $\ell s$ splitting,
size of the $\ell s$ splitting should not change much
even when the density-dependence is taken into account.
Therefore the remaining parameter $w_1$ has been fitted
to the splitting of the $n0i$ orbits obtained with M3Y-P6
at $^{208}$Pb~\cite{ref:NI15}.
This M3Y-P6a interaction will be applied  to other $Z=\mathrm{magic}$ nuclei.
Note that all the parameters of M3Y-P6a
except in $v^{(\mathrm{LS})}$ and $v^{(\mathrm{LS}\rho)}$
are identical to those of M3Y-P6.

Since $v^{(\mathrm{LS}\rho)}$ in Eq.~(\ref{eq:DDLS}) has zero range,
its contribution to the total energy can be represented
by a functional of local currents.
The particle-hole terms, which appear in the Hartree-Fock (HF) regime,
are summarized as
\begin{eqnarray}
E_{ph}^{(\mathrm{LS}\rho)} &=& \frac{1}{4}\int d^3r\,D[\rho(\mathbf{r})]\,
 \Big\{\rho(\mathbf{r})\,\nabla\cdot\mathbf{J}(\mathbf{r})
 + \sum_{\tau=p,n} \rho_\tau(\mathbf{r})\,\nabla\cdot\mathbf{J}_\tau(\mathbf{r})
 \nonumber\\
&&\qquad\qquad\qquad +i\,\mathbf{J}(\mathbf{r})\cdot\mathbf{j}^\ast(\mathbf{r})
 +i \sum_{\tau=p,n} \mathbf{J}_\tau(\mathbf{r})\cdot\mathbf{j}^\ast_\tau(\mathbf{r})
 \nonumber\\
&&\qquad\qquad\qquad -i\,\mathbf{J}^\ast(\mathbf{r})\cdot\mathbf{j}(\mathbf{r})
 -i \sum_{\tau=p,n} \mathbf{J}^\ast_\tau(\mathbf{r})\cdot\mathbf{j}_\tau(\mathbf{r})
 \nonumber\\
&&\qquad\qquad\qquad -\,\mathbf{Q}(\mathbf{r})\cdot\boldsymbol{\sigma}(\mathbf{r})
 - \sum_{\tau=p,n} \mathbf{Q}_\tau(\mathbf{r})\cdot\boldsymbol{\sigma}_\tau(\mathbf{r})
 \Big\} \,, \nonumber\\
 \label{eq:Eph_LSrho}
\end{eqnarray}
where the density $\rho(\mathbf{r})$ and the other local currents
are defined by
\begin{eqnarray}
 \rho(\mathbf{r})=\sum_{\tau=p,n} \rho_\tau(\mathbf{r})\,,&\quad&
 \rho_\tau(\mathbf{r})=\sum_{\alpha,\beta\in\tau}\varrho_{\alpha\beta}\,
 \phi_\beta^\dagger(\mathbf{r})\,\phi_\alpha(\mathbf{r})\,,
 \nonumber\\
 \mathbf{j}(\mathbf{r})=\sum_{\tau=p,n} \mathbf{j}_\tau(\mathbf{r})\,,&\quad&
 \mathbf{j}_\tau(\mathbf{r})=-i\sum_{\alpha,\beta\in\tau}\varrho_{\alpha\beta}\,
 \phi_\beta^\dagger(\mathbf{r})\,\nabla\phi_\alpha(\mathbf{r})\,,
 \nonumber\\
 \mathbf{Q}(\mathbf{r})=\sum_{\tau=p,n} \mathbf{Q}_\tau(\mathbf{r})\,,&\quad&
 \mathbf{Q}_\tau(\mathbf{r})=i\sum_{\alpha,\beta\in\tau}\varrho_{\alpha\beta}\,
 \nabla\phi_\beta^\dagger(\mathbf{r})\times\nabla\phi_\alpha(\mathbf{r})\,,
 \nonumber\\
 \boldsymbol{\sigma}(\mathbf{r})
 =\sum_{\tau=p,n} \boldsymbol{\sigma}_\tau(\mathbf{r})\,,&\quad&
 \boldsymbol{\sigma}_\tau(\mathbf{r})=2\sum_{\alpha,\beta\in\tau}\varrho_{\alpha\beta}\,
 \phi_\beta^\dagger(\mathbf{r})\,\mathbf{s}\,\phi_\alpha(\mathbf{r})\,,
 \nonumber\\
 \mathbf{J}(\mathbf{r})=\sum_{\tau=p,n} \mathbf{J}_\tau(\mathbf{r})\,,&\quad&
 \mathbf{J}_\tau(\mathbf{r})=2i\sum_{\alpha,\beta\in\tau}\varrho_{\alpha\beta}\,
 \phi_\beta^\dagger(\mathbf{r})\,\mathbf{s}\times\nabla\phi_\alpha(\mathbf{r})\,,
 \nonumber\\
 \label{eq:cur}
\end{eqnarray}
with the s.p. basis function $\phi_\alpha(\mathbf{r})$
and the one-body density matrix
$\varrho_{\alpha\beta}=\langle\Phi|a_\beta^\dagger a_\alpha|\Phi\rangle$
for the HF or HFB state $|\Phi\rangle$.
It should be noticed that, because of the presence of $D[\rho]$,
integration by parts does not simplify
the expression of $E_{ph}^{(\mathrm{LS}\rho)}$.

Under the spherical symmetry,
$\mathbf{Q}_\tau(\mathbf{r})=\boldsymbol{\sigma}_\tau(\mathbf{r})=0$,
$i[\mathbf{j}_\tau(\mathbf{r})-\mathbf{j}^\ast_\tau(\mathbf{r})]
=\nabla\rho_\tau(r)$
and $\mathbf{J}_\tau(\mathbf{r})=\mathbf{J}^\ast_\tau(\mathbf{r})
=(\mathbf{r}/r^2)\sum_{\alpha\beta\in\tau}\varrho_{\alpha\beta}\,
\phi_\beta^\dagger(\mathbf{r})\,(2\mbox{\boldmath$\ell$}\cdot\mathbf{s})\,
\phi_\alpha(\mathbf{r})$
with $\mbox{\boldmath$\ell$}=\mathbf{r}\times\mathbf{p}$
($\phi_\alpha$, $\phi_\beta$ are postulated to be spherical bases).
$E_{ph}^{(\mathrm{LS}\rho)}$ yields the $\ell s$ potential as follows~\cite{ls-corr}:
\begin{equation}
-\frac{1}{r}\bigg[\,D[\rho(r)]\,
 \frac{d}{dr}\Big(\rho(r)+\rho_\tau(r)\Big)
+ \frac{1}{2}\,\frac{\delta D}{\delta\rho}[\rho(r)]\,
 \Big(\rho(r)+\rho_\tau(r)\Big)\,\frac{d\rho(r)}{dr}\bigg]\,
 \mbox{\boldmath$\ell$}\cdot\mathbf{s}\,.\quad(\tau=p,n) \label{eq:lspot}
\end{equation}

The same Hamiltonian is applied to the pairing channels
in the HFB calculations,
including the two-body term of $H_\mathrm{c.m.}$.
Since $D[\rho(\mathbf{r})]$ does not contain the pairing tensor,
contribution of $v^{(\mathrm{LS}\rho)}$ to the pair energy
is represented analogously to that given in Ref.~\cite{ref:DFT84}.

The semi-realistic M3Y-P6a~\cite{ref:NI15} interaction is primarily used
in this work,
and its results are compared with those of M3Y-P6~\cite{ref:Nak13}
to exhibit effects of the density-dependence in the LS channel.
As mentioned above, the strength parameter $w_1$ in M3Y-P6a has been determined
so as to equate the $n0i$ splitting at $^{208}$Pb to its counterpart in M3Y-P6.
This makes influence on other $\ell s$ splitting insignificant as well.
For instance, the $n0d$ splitting at $^{40}$Ca decreases merely by $4\,\%$
if replacing M3Y-P6 by M3Y-P6a.
The equal-filling approximation~\cite{ref:EFA} is applied to odd-$N$ nuclei,
whose ground states have one quasiparticle.
It should be noted that the results of M3Y-P6 are qualitatively similar
to those of the other interactions
that lack density-dependence in the LS channels.
To show it,
the Gogny-D1M interaction~\cite{ref:D1M} will be taken as an example,
which has successfully been applied to describing structure of many nuclei.
While calculations beyond the spherical HFB have been implemented
with D1M and other globally fitted interactions,
qualitative improvement over the spherical HFB regime
has not been reported so far on the isotope shifts
of the nuclei under investigation, to my best knowledge.

\section{Results and discussions\label{sec:result}}
The frequency difference of corresponding atomic deexcitations among isotopes
is converted to the difference in mean-square (m.s.) charge radii of nuclei.
Hence the isotope shifts are expressed by the m.s. charge radius
of a certain nuclide relative to that of a reference nuclide
with equal atomic number $Z$.
The m.s. charge radius of the $^A Z$ nucleus
is given by
\begin{equation} \langle r^2\rangle_c(\mbox{$^A Z$})
= \langle r^2\rangle_p(\mbox{$^A Z$}) + \langle r^2\rangle_c(p)\,,
\label{eq:chrad}
\end{equation}
where $\langle r^2\rangle_p(\mbox{$^A Z$})$
represents the m.s. radius for point-proton distribution in $^A Z$
and $\langle r^2\rangle_c(p)$ the charge m.s. radius of a single proton.
By subtracting the c.m. contribution,
$\langle r^2\rangle_p(\mbox{$^A Z$})$ is obtained as
\begin{equation}
 \langle r^2\rangle_p(\mbox{$^A Z$}) = \frac{1}{Z}
 \sum_{i\in p}\langle\Phi|(\mathbf{r}_i-\mathbf{R})^2|\Phi\rangle\,.
 \quad \Big(\mathbf{R}=\frac{1}{A}\sum_i \mathbf{r}_i\Big)
\end{equation}
Because $\langle r^2\rangle_c(p)$ is canceled out,
the isotope shift is denoted
by $\mathit{\Delta}\langle r^2\rangle_p(\mbox{$^A Z$})
=\langle r^2\rangle_p(\mbox{$^A Z$})-\langle r^2\rangle_p(\mbox{$^{A_0} Z$})$
in this paper,
with $^{A_0} Z$ taken as a reference.

\subsection{Pb isotopes, revisited}
It is customary to define the isotope shift of the Pb nuclei
by adopting $^{208}$Pb as a reference,
$\mathit{\Delta}\langle r^2\rangle_p(\mbox{$^A$Pb})
=\langle r^2\rangle_p(\mbox{$^A$Pb})-\langle r^2\rangle_p(\mbox{$^{208}$Pb})$.
As has been shown by comparing the M3Y-P6 and M3Y-P6a results
in Ref.~\cite{ref:NI15},
the isotope shifts of the Pb nuclei can be improved
with the density-dependent LS interaction.

\begin{figure}
\includegraphics[scale=0.45]{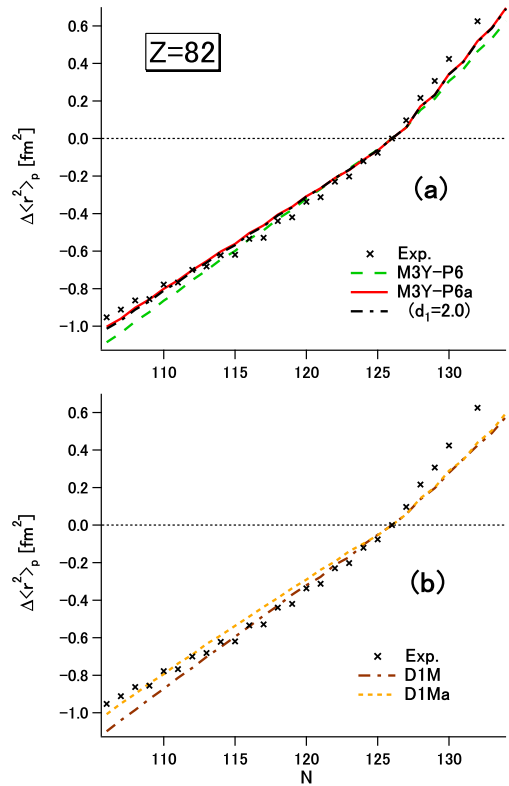}
\caption{(Color online) Isotope shifts of the Pb nuclei
$\mathit{\Delta}\langle r^2\rangle_p(\mbox{$^A$Pb})$
obtained from the HFB calculations.
(a) Comparison among the results with M3Y-P6 (green dashed line),
M3Y-P6a (red solid line) and its variant of $d_1=2.0\,\mathrm{fm}^3$
(black dot-dashed line).
(b) Comparison between the results with D1M (brown dot-dashed line)
and D1Ma (orange short-dashed line).
Experimental data quoted from Ref.~\cite{ref:Ang13} are shown by crosses
for reference.
\label{fig:Pb_drp}}
\end{figure}

It is exhibited in Fig.~\ref{fig:Pb_drp}(a)
that the $d_1$ parameter in Eq.~(\ref{eq:DinLS}),
which has been presumed to be $1.0\,\mathrm{fm}^3$ in M3Y-P6a,
hardly influences the results.
If $d_1$ is assumed to be $2.0\,\mathrm{fm}^3$,
$w_1=842\,\mathrm{MeV}\cdot\mathrm{fm}^8$ follows
by equating the $n0i$ splitting to that with M3Y-P6 at $^{208}$Pb,
instead of $742\,\mathrm{MeV}\cdot\mathrm{fm}^8$ in M3Y-P6a.
Notice that $D[\rho_0]$ values are close to each other.
The results with $d_1=2.0\,\mathrm{fm}^3$ are shown
by the black dot-dashed line.
Comparison to the results with M3Y-P6a
confirms insensitivity to $d_1$.

The density-dependent LS term in the effective interaction
may produce different behavior in the isotope shift
unless it is too weak,
as typically represented by the kink at $N=126$
in $\mathit{\Delta}\langle r^2\rangle_p(\mbox{$^A$Pb})$.
This is not constrained to the M3Y-P6 case,
although it depends on the interaction to certain degree
how strong its effects are.
It is here illustrated by D1M.
The LS channel of the D1M interaction is obtained
if $D[\rho]$ in $v^{(\mathrm{LS}\rho)}$ is replaced by a constant $w_0$.
The strength of $v^{(\mathrm{LS})}$ (\textit{i.e.} $w_0$)
was determined in a fully phenomenological manner.
When $v^{(\mathrm{LS}\rho)}$ is introduced, it is reasonable to reduce $w_0$
so as to keep size of the $\ell s$ splitting,
which leaves the ratio $w_1\rho_0/w_0$ almost arbitrary.
To see qualitative effects of $v^{(\mathrm{LS}\rho)}$,
$w_0$ is reduced to $30\,\%$ of its original value
and $w_1$ is adjusted by equating the $n0i$ splitting
to that of the original D1M,
with taking $d_1=1.0\,\mathrm{fm}^3$.
This interaction is called D1Ma in the present article.
The results of D1Ma on the isotope shifts of the Pb nuclei
are compared to the D1M results in Fig.~\ref{fig:Pb_drp}(b).
The experimental data are shown for reference,
although comparison to them is not the point here.
As has been pointed out~\cite{ref:RF95,ref:GSR13},
the slope of the isotope shifts becomes steeper in $N>126$ than in $N\leq 126$
because of the occupation on $n0i_{11/2}$.
With D1M (D1Ma), $n1g_{9/2}$ lies lower than $n0i_{11/2}$
by $1.4\,\mathrm{MeV}$ ($1.2\,\mathrm{MeV}$) in $^{208}$Pb.
This energy difference suppresses occupation on $n0i_{11/2}$ near $^{208}$Pb,
and therefore the effect of the density-dependent LS interaction
is hindered in $N>126$, though present.
However, because the $n0i_{13/2}$ function shrinks,
the isotope shift varies more slowly with D1Ma than with D1M in $N\leq 126$,
producing a visible kink at $N=126$.

\subsection{Ca isotopes}
\begin{figure}
\includegraphics[scale=0.45]{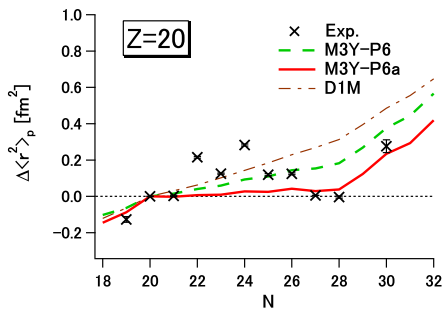}
\caption{(Color online) Isotope shifts of the Ca nuclei
$\mathit{\Delta}\langle r^2\rangle_p(\mbox{$^A$Ca})$,
obtained from the HFB calculations with M3Y-P6a (red solid line),
in comparison to those with M3Y-P6 (green dashed line)
and D1M (thin brown dot-dashed line).
Experimental data are quoted from Ref.~\cite{ref:Ang13} (crosses).
\label{fig:Ca_drp}}
\end{figure}

The isotope shifts of the Ca (\textit{i.e.} $Z=20$) nuclei
are defined by taking $^{40}$Ca as a reference nuclide,
$\mathit{\Delta}\langle r^2\rangle_p(\mbox{$^A$Ca})
=\langle r^2\rangle_p(\mbox{$^A$Ca})-\langle r^2\rangle_p(\mbox{$^{40}$Ca})$.
The HFB results are depicted and compared with the data
in Fig.~\ref{fig:Ca_drp}.
It is commented that the weak instabilities
against the octupole~\cite{ref:Nak13,ref:Nak08b}
and the pairing~\cite{ref:NS14} correlations in $^{40}$Ca
are lifted with M3Y-P6a.
The inversion of the $p0d_{3/2}$ and $p1s_{1/2}$ levels from $^{40}$Ca to $^{48}$Ca
is reproduced with M3Y-P6a
as well as with the previous interactions~\cite{ref:NSM13},
in which $v^{(\mathrm{TN})}$ plays a significant role.

As doubly magic nuclei,
both $^{40}$Ca and $^{48}$Ca are expected to be well described
within the HF regime.
Although it has been known that their charge radii are close to each other,
this property has been difficult to be reproduced
by self-consistent nuclear structure calculations so far.
It is interesting to see
$\mathit{\Delta}\langle r^2\rangle_p(\mbox{$^{48}$Ca}) \approx 0$,
\textit{i.e.} $\langle r^2\rangle_p(\mbox{$^{40}$Ca}) \approx
\langle r^2\rangle_p(\mbox{$^{48}$Ca})$, in the M3Y-P6a result.
Attraction from the neutrons occupying $0f_{7/2}$
determines $\mathit{\Delta}\langle r^2\rangle_p(\mbox{$^{48}$Ca})$.
As illustrated for $n0i$ orbits in Fig.~1 of Ref.~\cite{ref:NI15},
the density-dependence in the LS channel
tends to shrink the $j=\ell+1/2$ orbits while extends the $j=\ell-1/2$ orbits.
Therefore the m.s. radius of neutrons occupying the $0f_{7/2}$ orbit
comes smaller,
decreasing by $0.23\,\mathrm{fm}^2$ at $^{40}$Ca
when M3Y-P6 is switched to M3Y-P6a.
Moreover, the m.s. radius of $n0f_{7/2}$ increases from $^{40}$Ca to $^{48}$Ca
only by $0.035\,\mathrm{fm}^2$ with M3Y-P6a,
in contrast to $0.365\,\mathrm{fm}^2$ with M3Y-P6.
This is interpreted as the neutrons on $0f_{7/2}$ feel mutual attraction
and the density-dependent LS channel blocks them to distribute more broadly,
which further suppresses $\langle r^2\rangle_p$.
These effects lead to the result
$\mathit{\Delta}\langle r^2\rangle_p(\mbox{$^{48}$Ca}) \approx 0$.
Thus it may be possible to ascribe
the very small difference in the charge radii between $^{40}$Ca and $^{48}$Ca
to an effect of the $3N$ interaction.

Though not shown to keep the figure visible,
the density-dependence in the LS channel makes
$\mathit{\Delta}\langle r^2\rangle_p(\mbox{$^{48}$Ca})$ smaller
also when switching D1M to D1Ma,
confirming the qualitative effect of the density-dependence.
In quantitative respect, however,
$\mathit{\Delta}\langle r^2\rangle_p(\mbox{$^{48}$Ca})$ with D1Ma
is close to that with M3Y-P6,
mainly because D1M gives almost twice larger
$\mathit{\Delta}\langle r^2\rangle_p(\mbox{$^{48}$Ca})$ than M3Y-P6.

The isotope shifts of $^{42-46}$Ca
show sizable deviation from those of $^{40,48}$Ca.
Since $\langle r^2\rangle_p(\mbox{$^A$Ca})$ varies almost linearly
from $^{40}$Ca to $^{48}$Ca in the spherical HFB calculations,
none of the interactions including M3Y-P6a reproduce the observed $N$-dependence
of $\mathit{\Delta}\langle r^2\rangle_p(\mbox{$^A$Ca})$ in $N=22-26$.
This discrepancy may be ascribed to effects beyond MF,
including influence of the $\alpha$-clustering.
On the contrary,
the rising of $\mathit{\Delta}\langle r^2\rangle_p(\mbox{$^A$Ca})$
from $^{48}$Ca to $^{50}$Ca in the M3Y-P6a result
is in good agreement with the data,
while not so good in the M3Y-P6 and D1M results.
This also supports presence of the density-dependent LS interaction.

\subsection{Ni isotopes}
\begin{figure}
\includegraphics[scale=0.45]{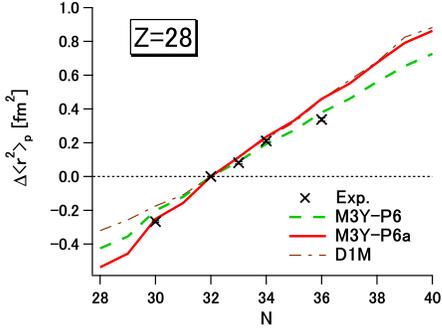}
\caption{(Color online) Isotope shifts of the Ni nuclei
$\mathit{\Delta}\langle r^2\rangle_p(\mbox{$^A$Ni})$.
See Fig.~\protect\ref{fig:Ca_drp} for conventions.
\label{fig:Ni_drp}}
\end{figure}

For the Ni (\textit{i.e.} $Z=28$) isotopes
$^{60}$Ni is used as a reference,
defining isotope shifts as $\mathit{\Delta}\langle r^2\rangle_p(\mbox{$^A$Ni})
=\langle r^2\rangle_p(\mbox{$^A$Ni})-\langle r^2\rangle_p(\mbox{$^{60}$Ni})$.
The results are depicted in Fig.~\ref{fig:Ni_drp}.

Steeper slope is obtained
for $\mathit{\Delta}\langle r^2\rangle_p(\mbox{$^A$Ni})$
with M3Y-P6a than with M3Y-P6 in $28\leq N\leq 40$.
This is attributed to contribution of neutrons occupying $n0f_{5/2}$,
which distribute more broadly
when the density-dependence is incorporated in the LS channel.
Compared to M3Y-P6,
M3Y-P6a gives slightly better agreement with the measured value for $^{58}$Ni
while slightly worse for $^{64}$Ni.
Since both of the M3Y-P6 and M3Y-P6a results are in reasonable agreement
with the existing data,
it is not obvious to tell whether the density-dependent LS channel
improves the theoretical results
on $\mathit{\Delta}\langle r^2\rangle_p(\mbox{$^A$Ni})$ or not.
It is desired to measure neutron-rich or neutron-deficient isotopes,
\textit{e.g.} $^{56}$Ni.

\subsection{Sn isotopes}
\begin{figure}
\includegraphics[scale=0.45]{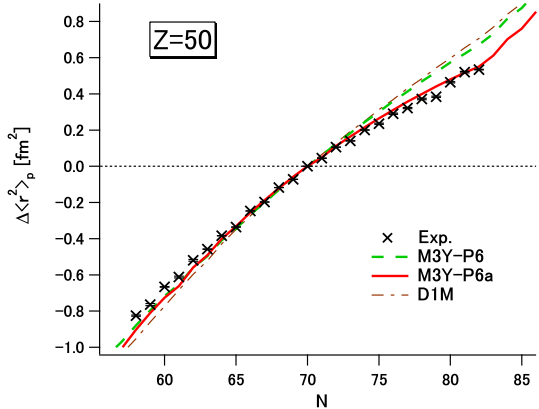}
\caption{(Color online) Isotope shifts of the Sn nuclei
$\mathit{\Delta}\langle r^2\rangle_p(\mbox{$^A$Sn})$.
See Fig.~\protect\ref{fig:Ca_drp} for conventions.
\label{fig:Sn_drp}}
\end{figure}

For the Sn (\textit{i.e.} $Z=50$) isotopes
$^{120}$Sn is adopted as a reference nuclide,
by which isotope shifts are defined
as $\mathit{\Delta}\langle r^2\rangle_p(\mbox{$^A$Sn})
=\langle r^2\rangle_p(\mbox{$^A$Sn})-\langle r^2\rangle_p(\mbox{$^{120}$Sn})$.
The results are presented in Fig.~\ref{fig:Sn_drp}.

It is found that $\mathit{\Delta}\langle r^2\rangle_p(\mbox{$^A$Sn})$
is well described with M3Y-P6a in a long chain of the Sn isotopes,
owing to the density-dependence in the LS channel.
The density-dependence gives narrower distribution of neutrons
occupying the $0h_{11/2}$ orbit,
whose attraction reduces $\mathit{\Delta}\langle r^2\rangle_p(\mbox{$^A$Sn})$
in $N>70$.
The same trend is obtained with D1Ma.

It is remarked that a kink is predicted at $N=82$ with the M3Y-P6a interaction.
Origin of this kink is accounted for analogously to the Pb case.
Attraction from neutrons occupying $0h_{9/2}$ contributes
to the broader proton distribution.
While occupation probability on $n0h_{9/2}$ is negligibly small in $N\leq 82$,
$n0h_{9/2}$ is partially occupied in $N>82$ owing to the pairing correlation,
although the lowest s.p. level above $N=82$ is $n1f_{7/2}$. 
The relatively large m.s. radius of $n0h_{9/2}$,
for which the density-dependence in the LS channel is responsible,
produces the kink at $N=82$.
The reduction of the radii in $N<82$,
an effect of the density-dependent LS channel on $n0h_{11/2}$,
makes the kink more conspicuous.
This kink in $\mathit{\Delta}\langle r^2\rangle_p(\mbox{$^A$Sn})$ at $N=82$
is a clear contrast to the results of M3Y-P6 and D1M
which have no density-dependent LS channel.
Thus the density-dependence in the LS channel is essential to the kink.
Future measurements with respect to this kink
(\textit{i.e.} $\mathit{\Delta}\langle r^2\rangle_p(\mbox{$^A$Sn})$
beyond $N=82$) will be intriguing,
which could be a touchstone of the $3N$ LS effects.

\subsection{Absolute values of charge radii and neutron-skin thickness}

Although the isotope shifts are the main subject of this paper,
it deserves discussing absolute values of the charge radii
of the reference nuclei as well.
While the finite-size effects of protons are canceled out in the isotope shifts,
absolute values of the charge radii are affected by the ambiguity
in the charge radius of an isolated proton.
A recent compilation~\cite{ref:PDG14} gives two values,
$\langle r^2\rangle_c(p)=0.84087\pm 0.00039$ by $\mu p$ Lamb shift
and $0.8775\pm 0.0051$ by $e p$ CODATA.
The spherical HFB results on the root-m.s. charge radii
$\sqrt{\langle r^2\rangle_c(\mbox{$^A Z$})}$ with M3Y-P6 and M3Y-P6a
are tabulated in Table~\ref{tab:abs-value},
obtained from Eq.~(\ref{eq:chrad})
by taking both of the $\mu p$ and the $e p$ values
for $\langle r^2\rangle_c(p)$,
in comparison to the experimental data~\cite{ref:Ang13}.

\begin{table}
~\vspace*{-2.3cm}
\begin{center}
\caption{Root-m.s. charge radii of the reference nuclei.
Spherical HFB results with M3Y-P6 and M3Y-P6a
are compared with experimental data~\cite{ref:Ang13}.
$(\mu p)$ and $(ep)$ corresponds to the data on $\langle r^2\rangle_c(p)$.
\label{tab:abs-value}}
\begin{tabular}{ccrrr@{$\pm$}l}
\hline\hline
nuclide && ~~M3Y-P6 & ~~M3Y-P6a & \multicolumn{2}{c}{Exp.} \\
 \hline
$^{16}$O &($\mu p$)& $2.732$~~& $2.725$~~&~~$2.6991$&$0.0052$\\
 &($e p$)& $2.743$~~& $2.737$~~&\multicolumn{2}{c}{~}\\
$^{40}$Ca &($\mu p$)& $3.491$~~& $3.487$~~&~~$3.4776$&$0.0019$\\
 &($e p$)& $3.500$~~& $3.496$~~&\multicolumn{2}{c}{~}\\
$^{60}$Ni &($\mu p$)& $3.825$~~&$3.779$~~&~~$3.8118$&$0.0016$\\
 &($e p$)& $3.834$~~& $3.787$~~&\multicolumn{2}{c}{~}\\
$^{120}$Sn &($\mu p$)& $4.653$~~& $4.627$~~&~~$4.6519$&$0.0021$\\
 &($e p$)& $4.659$~~& $4.634$~~&\multicolumn{2}{c}{~}\\
$^{208}$Pb &($\mu p$)& $5.493$~~& $5.464$~~&~~$5.5012$&$0.0013$\\
 &($e p$)& $5.499$~~& $5.470$~~&\multicolumn{2}{c}{~}\\
\hline\hline
\end{tabular}
\end{center}
\end{table}

It is found via comparison of the M3Y-P6a results to the M3Y-P6 ones
that the density-dependence in the LS channel tends to reduce the radii.
Whereas this effect is weak in $^{16}$O and $^{40}$Ca
which are so-called LS-closed nuclei,
it is stronger in $^{60}$Ni, $^{120}$Sn and $^{208}$Pb,
with $0.02-0.05\,\mathrm{fm}$ difference between the two interactions.
It is reasonable that the effect is minimal in the LS-closed nuclei
in which all the $\ell s$ partners (the $j=\ell\pm 1/2$ orbitals) are filled,
in contrast to the $jj$-closed nuclei
in which there is a pair of an occupied $j=\ell+1/2$
and unoccupied $j=\ell-1/2$ orbitals.
Since the parameters of the central channels
have been more or less fitted to the measured radius of $^{208}$Pb,
the charge radius is well reproduced with M3Y-P6 for this nucleus,
and so for $^{120}$Sn.
This indicates that M3Y-P6a slightly underestimates the charge radii
of $^{120}$Sn and $^{208}$Pb,
although it gives values closer to the measured ones than M3Y-P6
for $^{16}$O and $^{40}$Ca.
As the MF approaches are expected to be the more appropriate
for the heavier nuclei,
there may be a room to readjust the parameters in the central channels.
Such readjustment could influence the saturation density by $1\,\%$.

The neutron-skin thickness,
which is usually represented
by the difference of the neutron and proton root-m.s. radii
$\sqrt{\langle r^2\rangle_n}-\sqrt{\langle r^2\rangle_p}$ in a single nucleus,
attracts interests because it correlates to variation of the symmetry energy
for increasing density~\cite{ref:CR09}.
Because of their neutron excess and of experimental accessibility,
the neutron-skin thickness is frequently argued
for $^{48}$Ca, $^{68}$Ni, $^{132}$Sn and $^{208}$Pb.
The density-dependence in the LS channel hardly influences
this quantity at $^{208}$Pb and at $^{132}$Sn;
difference between M3Y-P6 and M3Y-P6a is $0.001\,\mathrm{fm}$ or less.
On the other hand, 
$\sqrt{\langle r^2\rangle_n}-\sqrt{\langle r^2\rangle_p}$ is larger
by $0.011\,\mathrm{fm}$ with M3Y-P6a than with M3Y-P6 at $^{68}$Ni
and smaller by $0.014\,\mathrm{fm}$ at $^{48}$Ca.
This is because the effects of the density-dependence
tend to be canceled if both of the $\ell s$ partners are occupied.
Though not very strong,
influence of the density-dependence in the LS channel
cannot be discarded for certain nuclides
in order to argue their neutron-skin thickness to $0.01\,\mathrm{fm}$ accuracy.

\section{Summary\label{sec:summary}}
Effects of the density-dependent LS interaction,
which was suggested from the chiral $3N$ interaction,
have been extensively studied on the isotope shifts of the Ca, Ni and Sn nuclei.
Since the density-dependence makes the LS interaction stronger
in the nuclear interior relative to its exterior,
the s.p. wave functions of the $j=\ell+1/2$ orbits shrink
while those of the $j=\ell-1/2$ orbits become broader,
if the $\ell s$ splitting is maintained.
It was pointed out in the previous paper that, owing to this mechanism,
the kink in the isotope shifts of the Pb nuclei at $N=126$
can be better described by introducing the density-dependent LS channel.

The spherical HFB calculations have been implemented with M3Y-P6a,
which contains density-dependence in the LS channel
but is identical to M3Y-P6 in the other channels.
It is found that almost equal charge radii between $^{40}$Ca and $^{48}$Ca
have been described well,
although this property has been difficult to be reproduced
in the MF calculations so far.
Moreover, the isotope shifts of the Sn nuclei are in excellent agreement
with the available data.
A kink is predicted at $N=82$ in the isotope shifts of Sn,
in contrast to those obtained from interactions
with no density-dependent LS channel.
Future experiments on the isotope shifts of Sn beyond $N=82$ are awaited,
as well as on $^{56}$Ni
for which predictions significantly depend on interactions.

Absolute values of the charge radii have also been argued.
The density-dependence in the LS channel reduces the absolute values,
amounting to about $1\,\%$ except for the LS-closed nuclei.
The density-dependence may influence the neutron-skin thickness
in $^{48}$Ca and $^{68}$Ni by about $0.01\,\mathrm{fm}$.
Both are small but not necessarily negligible.

\begin{acknowledgments}
%
The author is grateful to J. Margueron for discussions.
This work is financially supported in part
as Grant-in-Aid for Scientific Research on Innovative Areas,
No.~24105008, by The Ministry of Education, Culture, Sports, Science
and Technology, Japan,
and as Grant-in-Aid for Scientific Research (C), No.~25400245,
by Japan Society for the Promotion of Science.
Numerical calculations are performed on HITAC SR16000s
at Institute of Management and Information Technologies in Chiba University,
Information Technology Center in University of Tokyo,
Information Initiative Center in Hokkaido University,
and Yukawa Institute for Theoretical Physics in Kyoto University.
\end{acknowledgments}


\begin{thebibliography}{99}
\bibitem{ref:AB81} K. And\={o} and H. Band\={o},
 Prog. Theor. Phys. \textbf{66}, 227 (1981).
\bibitem{ref:EGM05} E. Epelbaum, W. Gl\"{o}ckle and U.-G. Mei\ss ner,
 Nucl. Phys. A \textbf{747}, 362 (2005).
\bibitem{ref:Kai03} N. Kaiser, Phys. Rev. C \textbf{68}, 054001 (2003).
\bibitem{ref:Koh12} M. Kohno, Phys. Rev. C \textbf{86}, 061301(R) (2012).
\bibitem{ref:Koh13} M. Kohno, Phys. Rev. C \textbf{88}, 064005 (2013).
\bibitem{ref:AHS87} P. Aufmuth, K. Heilig and A. Steudel,
 At. Data Nucl. Data Tables \textbf{37}, 455 (1987).
\bibitem{ref:Ang04} I. Angeli,
 At. Data Nucl. Data Tables \textbf{87}, 185 (2004).
\bibitem{ref:SLR94} M.M. Sharma, G. Lalazissis and P. Ring,
 Phys. Lett. B \textbf{317}, 9 (1994).
\bibitem{ref:SLKR95} M.M. Sharma, G. Lalazissis, J. K\"{o}nig and P. Ring,
 Phys. Rev. Lett. \textbf{74}, 3744 (1995).
\bibitem{ref:RF95} P.-G. Reinhard and H. Flocard,
 Nucl. Phys. A \textbf{584}, 467 (1995).
\bibitem{ref:TI} R.B. Firestone \textit{et al.},
\textit{Table of Isotopes}, 8th edition
(John Wiley \& Sons, New York, 1996).
\bibitem{ref:NI15} H. Nakada and T. Inakura,
 Phys. Rev. C \textbf{91}, 021302(R) (2015).
\bibitem{ref:PF94} J.M. Pearson and M. Farine,
 Phys. Rev. C \textbf{50}, 185 (1994).
\bibitem{ref:Gor15} S. Goriely, Nucl. Phys. A \textbf{933}, 68 (2015).
\bibitem{ref:GEM} E. Hiyama, Y. Kino and M. Kamimura,
 Prog. Part. Nucl. Phys. \textbf{51} (2003) 223.
\bibitem{ref:Nak08} H. Nakada, Nucl. Phys. A \textbf{808}, 47 (2008).
\bibitem{ref:Nak08b} H. Nakada, Phys. Rev. C \textbf{78}, 054301 (2008).
\bibitem{ref:Nak10} H. Nakada, Phys. Rev. C \textbf{81}, 027301 (2010).
\bibitem{ref:Nak13} H. Nakada, Phys. Rev. C \textbf{87}, 014336 (2013).
\bibitem{ref:M3Y} G. Bertsch, J. Borysowicz, H. McManus and W.G. Love,
 Nucl. Phys. A \textbf{284}, 399 (1977).
\bibitem{ref:M3Y-P} N. Anantaraman, H. Toki and G.F. Bertsch,
 Nucl. Phys. A \textbf{398}, 269 (1983).
\bibitem{ref:Nak03} H. Nakada, Phys. Rev. C \textbf{68}, 014316 (2003).
\bibitem{ref:Nak10b} H. Nakada, Phys. Rev. C \textbf{81}, 051302(R) (2010).
\bibitem{ref:NSM13} H. Nakada, K. Sugiura and J. Margueron,
 Phys. Rev. C \textbf{87}, 067305 (2013).
\bibitem{ref:NS14} H. Nakada and K. Sugiura,
 Prog. Theor. Exp. Phys. \textbf{2014}, 033D02.
\bibitem{unit-corr}
 The unit of $t_\rho^{(\mathrm{SE})}$ ($t_\rho^{(\mathrm{TE})}$)
 in Table~I of Ref.~\cite{ref:Nak13} should be corrected
 to $\mathrm{MeV}\,\mathrm{fm}^6$ ($\mathrm{MeV}\,\mathrm{fm}^4$).
\bibitem{ls-corr}
 There was an error in Eq.~(5) in Ref.~\protect\cite{ref:NI15},
 which should be multiplied by the factor 2.
\bibitem{ref:DFT84} J. Dobaczewski, H. Flocard and J. Treiner,
 Nucl. Phys. A \textbf{422}, 103 (1984).
\bibitem{ref:EFA} S. Perez-Martin and L.M. Robledo,
 Phys. Rev. C \textbf{78}, 014304 (2008);
N. Schunck \textit{et al.},
 Phys. Rev. C \textbf{81}, 024316 (2010).
\bibitem{ref:D1M} S. Goriely, S. Hilaire, M. Girod and S. P\`{e}ru,
 Phys. Rev. Lett. \textbf{102}, 242501 (2009).
\bibitem{ref:Ang13} I. Angeli and K.P. Marinova,
 At. Data Nucl. Data Tables \textbf{99}, 69 (2013).
\bibitem{ref:GSR13} P.M. Goddard, P.D. Stevenson and A. Rios
 Phys. Rev. Lett. \textbf{110}, 032503 (2013).
\bibitem{ref:PDG14} K.A. Olive \textit{et al.} (Particle Data Group),
 Chin. Phys. C \textbf{38}, 090001 (2014).
\bibitem{ref:CR09} M. Centelles, X. Roca-Maza, X. Vi\~{n}as and M. Warda,
 Phys. Rev. Lett. \textbf{102}, 122502 (2009).
\end{thebibliography}

\end{document}